\documentclass[a4paper,11pt]{article}
\usepackage{pos}

\usepackage{tikz}
\usepackage{epsfig}
\usepackage{bbm}
\usepackage{bbold}
\usepackage{graphicx}
\usepackage{hyperref}
\usepackage{graphics,graphpap}
\usepackage{yfonts}
\usepackage{mathrsfs}
\usepackage{float}
\usepackage{hyperref}
\usepackage{subcaption}
 \usepackage{feynmf}
 \usepackage{slashed}

\begin{document}

\def\a{\alpha}
\def\b{\beta}
\def\c{\chi}
\def\d{\delta}
\def\e{\epsilon}
\def\f{\phi}
\def\g{\gamma}
\def\h{\eta}
\def\i{\iota}
\def\j{\psi}
\def\k{\kappa}
\def\la{\lambda}
\def\m{\mu}
\def\n{\nu}
\def\o{\omega}
\def\p{\pi}
\def\q{\theta}
\def\r{\rho}
\def\s{\sigma}
\def\t{\tau}
\def\u{\upsilon}
\def\x{\xi}
\def\z{\zeta}
\def\D{\Delta}
\def\F{\Phi}
\def\G{\Gamma}
\def\J{\Psi}
\def\L{\Lambda}
\def\O{\Omega}
\def\P{\Pi}
\def\Q{\Theta}
\def\S{\Sigma}
\def\U{\Upsilon}
\def\X{\Xi}

\def\ve{\varepsilon}
\def\vf{\varphi}
\def\vr{\varrho}
\def\vs{\varsigma}
\def\vq{\vartheta}

\def\dg{\dagger}                                     
\def\ddg{\ddagger}                                   
\def\wt#1{\widetilde{#1}}                    
\def\mt{\widetilde{m}_1}
\def\mti{\widetilde{m}_i}
\def\rt{\widetilde{r}_1}
\def\mtt{\widetilde{m}_2}
\def\mttt{\widetilde{m}_3}
\def\rtt{\widetilde{r}_2}
\def\mb{\overline{m}}
\def\VEV#1{\left\langle #1\right\rangle}        
\def\be{\begin{equation}}
\def\ee{\end{equation}}
\def\ds{\displaystyle}
\def\ra{\rightarrow}

\def\bea{\begin{eqnarray}}
\def\eea{\end{eqnarray}}
\def\NO{\nonumber}
\def\Bar#1{\overline{#1}}


\def\pl#1#2#3{Phys.~Lett.~{\bf B {#1}} ({#2}) #3}
\def\np#1#2#3{Nucl.~Phys.~{\bf B {#1}} ({#2}) #3}
\def\prl#1#2#3{Phys.~Rev.~Lett.~{\bf #1} ({#2}) #3}
\def\pr#1#2#3{Phys.~Rev.~{\bf D {#1}} ({#2}) #3}
\def\zp#1#2#3{Z.~Phys.~{\bf C {#1}} ({#2}) #3}
\def\cqg#1#2#3{Class.~and Quantum Grav.~{\bf {#1}} ({#2}) #3}
\def\cmp#1#2#3{Commun.~Math.~Phys.~{\bf {#1}} ({#2}) #3}
\def\jmp#1#2#3{J.~Math.~Phys.~{\bf {#1}} ({#2}) #3}
\def\ap#1#2#3{Ann.~of Phys.~{\bf {#1}} ({#2}) #3}
\def\prep#1#2#3{Phys.~Rep.~{\bf {#1}C} ({#2}) #3}
\def\ptp#1#2#3{Progr.~Theor.~Phys.~{\bf {#1}} ({#2}) #3}
\def\ijmp#1#2#3{Int.~J.~Mod.~Phys.~{\bf A {#1}} ({#2}) #3}
\def\mpl#1#2#3{Mod.~Phys.~Lett.~{\bf A {#1}} ({#2}) #3}
\def\nc#1#2#3{Nuovo Cim.~{\bf {#1}} ({#2}) #3}
\def\ibid#1#2#3{{\it ibid.}~{\bf {#1}} ({#2}) #3}

\title{Dark matter from sterile-sterile neutrino mixing}

\author[a]{Pasquale Di Bari}

\affiliation[a]{School of Physics and Astronomy, University of Southampton\\
  Southampton, SO17 1BJ, U.K.}


\emailAdd{P.Di-Bari@soton.ac.uk}

\abstract{A solution to the problem of the origin of matter in the universe can be reasonably searched within extensions of the standard model that also
explain neutrino masses and mixing. Models embedding the minimal seesaw mechanism can explain the observed matter-antimatter asymmetry of the 
universe via leptogenesis and dark matter via active-sterile neutrino mixing. In this case a keV lightest seesaw neutrino would play 
the role of warm dark matter particle. This traditional solution is now constrained by various cosmological observations. I will discuss the possibility that  a much heavier but yet metastable (dark) right-handed neutrino with mass in the $1\,{\rm TeV}$--$1 \, {\rm PeV}$ range can play the role of (cold) dark matter particle. The right abundance would be produced by the Higgs induced mixing with a seesaw right-handed neutrino (RHINO model), 
i.e., by sterile-sterile neutrino mixing.  Such a mixing would necessarily require a further extension of the minimal seesaw mechanism and can be described by a dimension-five effective operator.  The same mixing would also necessarily induce dark neutrino instability with lifetimes that can be much longer than the age of the universe and can escape current constraints from neutrino telescopes.  On the other hand, 
a contribution to very high energy neutrino flux  produced by dark neutrino decays could explain
an anomalous excess at 100 TeV energies confirmed recently by the IceCube collaboration. I will also discuss a simple UV completion where the mediator is given by a massive fermion. Intriguingly, it comes out that the favoured scale of new physics for RHINO to satisfy the dark matter
requirements coincides with the grand-unified scale: a RHINO miracle.}

\FullConference{%
  Corfu Summer Institute 2022 "School and Workshops on Elementary Particle Physics and Gravity",\\
  28 August - 1 October, 2022\\
  Corfu, Greece}

\maketitle

\section{Introduction}

The origin of matter in the universe is one of the greatest mystery in science \cite{origin}. Our observable universe is maximallly matter-antimatter
asymmetric, since observations do not find evidence of primordial antimatter.  The current baryonic matter abundance is then the the leftover matter-antimatter asymmetry after annihilations took place in the early universe. 

Combining {\em Planck} satellite results on CMB anisotropies
and baryon acoustic oscillation data, the {\em Planck} collaboration
finds for the baryonic contribution to the energy density parameter \cite{planck18}
\be\label{OB0}
\O_{B0}h^2 = 0.02242 \pm 0.00014 \,  ,
\ee
that translates into a total baryon-to-photon number ratio 
\be\label{etaB}
\eta_{B 0} \equiv {n_{B 0} - n_{\bar{B} 0}\over n_{\gamma 0}} \simeq {\O_{B0}\ ,\varepsilon_{\rm c 0}\over m_p \, n_{\gamma 0}}
\simeq 273.5\,\O_{B 0}h^2\,10^{-10} = (6.12 \pm 0.04)\times 10^{-10} \,  .
\ee
This matter-antimatter asymmetry cannot be explained within the standard model (SM) and, therefore, 
it is regarded as a strong motivation for the existence of new physics. 

At the same time cosmological observations point to the existence of a form of matter of non-standard nature referred as dark matter (DM). 
Its existence is necessary at high redshifts to explain observations of galactic rotation curves, dynamics of galaxy clusters and gravitational lensing in systems like the bullet cluster. 
It would indeed act as a kind of invisible fluid necessary to explain gravitational interactions in these systems. Moreover, DM had to be produced in the early universe, playing a role primordially in order to understand structure formation and CMB temperature anisotropies. All these observations favour an interpretation of this fluid in terms of the existence of new non-standard particle(s). This fluid had to be cold enough\footnote{This implies a free streaming (comoving) length  at the matter-radiation equality time not larger than about $0.1\,{\rm Mpc}$ \cite{Colin:2000dn}.} in order for numerical simulations to reproduce the observed large scale structure  of the universe. 

The cold DM contribution to the energy density parameter is today well determined both from dynamics of clusters of galaxies and from CMB anisotropies and from {\em Planck} +  BAO data it is found \cite{planck18}
\be\label{ODM}
\O_{{\rm DM} 0} h^2 = 0.11933 \pm 0.00091 \simeq 5 \, \O_{{\rm B} 0}h^2 \,  .
\ee
There is no SM particle with properties able to satisfy the DM requirements in order to explain the cosmological observations and for this
reason the DM puzzle is also regarded as a strong motivation for the existence of new physics. 

For long time the dominant solution has been represented by weakly interacting massive particles (WIMPs) with a relic abundance explained by
a traditional non-relativistic freeze-out (the WIMP paradigm). 
This was mainly supported by the observation that the observed DM abundance could be explained
for WIMP masses of order of the electroweak scale (so-called WIMP miracle). However, negative results from direct and indirect searches
require a relaxation of the assumptions of the WIMP miracle paradigm, so that WIMPs, though still not excluded as possible DM candidates, 
do not represent any more the standard solution to the DM puzzle and many other models have been proposed and investigated. 
The strong constraints imposed by direct and indirect searches require that the new physics necessary to provide a solution to the DM puzzle must lie (i) either at energy scales higher than those accessible at the LHC or (ii) involve sufficiently small interaction couplings 
or (iii) some combination of the two. The  solution I will discuss in this talk is indeed of this third kind since it 
relies both on very high energy scales and very small couplings.

The matter-antimatter asymmetry of the universe and the DM puzzles can be combined together
in the problem of the origin of  matter in the universe. It is indeed reasonable that their solution might 
stem from the same kind of new physics and it is, therefore, also reasonable to investigate extensions 
of the SM able to address these two problems jointly.

From neutrino oscillations we know that neutrinos are massive and mix so that neutrino physics 
also provides a strong phenomenological motivation to extend the SM.  
It is then quite well motivated to search for a solution to the origin of matter of the universe  
within extensions of the SM that incorporate neutrino masses and mixing. 

In this talk I will discuss the possibility that a right-handed (RH) neutrino in the TeV-PeV mass range can play the role of DM particle. This window is not typically considered but it is very interesting, since it is currently tested by IceCube, and more generally by neutrino telescopes. 


\section{Minimal type-I seesaw mechanism}

Adding $N$ RH neutrinos to the SM Lagrangian and a right-right Majorana mass term violating lepton number, 
one can add to the SM Lagrangian the terms ($\a =e,\mu,\tau$; $I =1,2,\dots, N$)
\be\label{Y+M}
- {\cal L}_{Y+M}^{\nu}=\overline{L_{\a}}\,h^{\nu}_{\a I}\,\nu_{R I}\, \widetilde{\Phi} \,   +
{1\over 2}\,\overline{\nu_{R \, I}^{\,c}}\,M_{I}\,\nu_{R \, I} + {\rm h.c.} \,  ,
\ee
written in a basis where charged lepton and Majorana mass matrices are diagonal. 
 After electroweak spontaneous symmetry breaking, we have that the neutrino
mass term can be written  as 
\be
-{\cal L}^{\nu}_{\rm m}= {1\over 2}\,
\left[(\overline{\nu_L},\overline{\nu_R^{c}})
\left(
\begin{array}{cc}
                0  & m_D  \\
               m_D^T &  M    \\
\end{array}\right)
\left(
\begin{array}{c}
       \nu_L^{c}  \\
       \nu_R  \\
\end{array}\right)
\right] + {\rm h.c.} \,  .
\ee
In the seesaw limit, for $M \gg m_D$, the spectrum of neutrino masses splits into a light set (ordinary neutrinos) with masses
given by the seesaw formula \cite{seesaw}
\be\label{seesawdiagonal}
{\rm diag}(m_1,m_2,m_3) =  U_L^{\nu \dagger}\,m_D\,\mbox{\large ${1\over M}$}\,m_D^T\, U_L^{\nu \star} \,  ,
\ee
where in the particular flavour basis we are taking the diagonalising unitary matrix $U$ can be identified 
with the leptonic mixing matrix, and into a heavy set (seesaw neutrinos) with masses $M_1 \leq  M_2 \leq \dots \leq M_N$. 
For definiteness I will consider the case $N=3$. 

If one assumes that the reheat temperature of the universe $T_{\rm RH}$ is not much lighter than the lightest 
heavy neutrino state $N_1$ with mass $M_1$, then the $N_1$ decays, and possibly also those of the other heavier seesaw neutrinos,
can generate a lepton asymmetry that can be partly converted into a baryon asymmetry by sphalerons if $T_{\rm RH} \gtrsim 132\, {\rm GeV}$ \cite{rummukainen}.  One has also to take into account that part of this asymmetry will be washed-out by inverse processes but it can be shown that the 
measured baryon-to-photon number ratio in Eq.~(\ref{etaB}) can be nicely reproduced: this is the essence of thermal leptogenesis \cite{fy}, 
a model of baryogenesis  that is a straightforward cosmological  application of the seesaw mechanism \cite{Blanchet:2012bk}.

It is clearly intriguing to think whether such a simple extension of the SM might also address the DM puzzle. 
If one writes the light and heavy mass eigenfields, that are Majorana fields, 
as a linear combination of left-handed (LH) and RH neutrinos, then light neutrinos are dominantly LH neutrinos and heavy neutrinos
are dominantly RH neutrinos. However, they also have subdominant RH (sterile) and LH (active) components mixing together 
and the mixing is described by $m_D\,M^{-1}$. If one of the RH neutrinos has sufficiently long lifetime and at the same time be produced by
such active-sterile neutrino mixing, then it could play the role of DM particle. 
There is a very attractive solution to such a simultaneous request \cite{dw}.  
If $M_1 \ll m_e$, then the $N_1$ lifetime is given by 
\be
\tau_1 \simeq 5 \times 10^{28}\,{\rm s}\,\left({M_1 \over {\rm keV}} \right)^{-5} \, \left({10^{-4} \over \theta}\right)^2  \,   ,
 \ee
 where $\theta \equiv \sum_\a \, |m_{D\a 1} / M_1 |^2$ is an effective active-sterile neutrino mixing angle.
One can see that $N_1$ can be long enough not only to satisfy the DM long-life condition but also to evade $X$-ray constraints, since it would
subdominantly radiatively decay. At the same time such a $\sim {\rm keV}$ seesaw neutrino would mainly mix with the lightest
ordinary neutrino mass eigenstate, one has specifically:
\bea
\nu_1 & = & \left[U^\dagger_{1\a}\nu_{L\a} + (U^\dagger_{1\a}\nu_{L\a})^c \right]  -  
\left[V^\nu_{L 1\a}\,\xi^\star_{\a 1} \,\nu_{R 1} + (V^\nu_{L 1\a}\,\xi^\star_{\a 1} \,\nu_{R1})^c\right]  \\
N_1  & =&  \left[ \nu_{R 1} + \nu_{R1}^c \right] + 
         \left[\xi^T_{1\a} \,\nu_{L \a} + (\xi^T_{1\a} \,\nu_{L \a})^c \right]  \,   .
\eea
It is this mixing that is actually responsible for the $N_1$-decays. At the same time it would also be responsible for the production
of a $N_1$-abundance given by
\be
\O_{N_1} h^2 \sim 0.1 \, \left({\theta \over 10^{-4} }\right)^2 \,  \left({M_1 \over {\rm keV}} \right)^{2} \,  ,
\ee
such that the measured DM abundance in Eq.~(\ref{ODM}) can  be reproduced 
for $\theta\sim 10^{-4}$ and $M_1 \sim {\rm keV}$ \cite{nuMSM}.   Interestingly, for keV masses, 
the lightest seesaw neutrino would behave as warm DM, implying that compared to cold DM there would be a reduced power in the large scale structure at scales corresponding to dwarf galaxies ($\sim 0.1\,{\rm Mpc}$ in comoving length). This would also
help in solving some potential issues in pure cold DM N-body simulations that seem to predict too
many satellite galaxies, in a galaxy like ours, compare to what is observed astronomically. At the same time warm DM
would also smooth the cusp profile in galaxies that is predicted by cold DM N-body simulations but that is
in tension with different observations.  

On the other hand, X-ray observations place an upper bound to the mass $M_1$, while galaxy distribution constraints require that the
$N_1$'s are not too warm DM particles and place a lower bound on $M_1$. These constraints have progressively closed down the allowed window 
ruling out such a minimal solution. However,  if a large pre-existing lepton asymmetry, $L \sim 10^{-5}$--$10^{-4}$, is present prior to the mixing, then the DM production is resonantly enhanced \cite{shifuller,Dolgov:2000ew} and this can reconcile the DM requirements with the $X$-ray observations.  
Moreover, it is very interesting that the observation of a new  $3.5\,{\rm keV}$ line in the X-ray observations of different 
clusters of galaxies \cite{bulbul,boyarsky} could be explained by such a mechanism for a decaying  
$7\,{\rm keV}$ sterile neutrino with a mixing angle $\theta \simeq 4 \times 10^{-6}$ and its 
relic abundance can explain the observed DM density parameter for a 
lepton asymmetry  at the resonance $L \simeq 4.6 \times 10^{-4}$ \cite{abazajian}.
However, authors of more recent observations of the 3.5 keV line
claimed to exclude an interpretation in terms of DM decays though this seems to be currently
controversially debated \cite{replies}. It is fair to conclude 
that more observations will be likely necessary to definitively test 
this exciting possibility, in particular the XRISM satellite should provide in 
the next years a final answer to the 3.5 keV anomaly \cite{XRISM}. 

Finally, let us comment that such a solution to the DM puzzle can also be combined with leptogenesis 
from the mixing of two heaviest seesaw neutrinos ($N_2$ and $N_3$ in our notation) with GeV masses \cite{ars,asaka} that would be also
responsible for the atmospheric and solar neutrino mass scales thanks to the seesaw formul. In this way one would 
realise a unified picture of neutrino masses, DM and leptogenesis, the so-called $\nu$MSM model.
However, such unified solution seems to rely on a very strong mass degeneracy of $N_2$ and $N_3$,
at the level of $(M_3-M_2)/(M_2+M_3) \simeq 10^{-16}$ \cite{ghiglieri}. The FASER experiment should 
be able to test the existence of GeV RH neutrinos during next years. These should be produced in B meson decays 
for masses $M_{2,3} \lesssim M_B \sim 3 \,{\rm GeV}$ \cite{faser}. 

\section{Heavy RH neutrino as DM}

For very heavy DM particles ($M_{\rm DM} \gtrsim 1\,{\rm TeV}$), just a tiny non-thermal abundance is sufficient to
reproduce the observed energy density parameter, since one has:
\be
N_{\rm DM} \sim 10^{-9}\,(\O_{{\rm DM} 0} h^2)\, N_{\gamma}^{\rm prod}\,{{\rm TeV}\over M_{\rm DM}} \sim 10^{-10}\,N_{\gamma}^{\rm prod}\,{{\rm TeV}\over M_{\rm DM}} \,  .
\ee
This simple observation is encouraging, since it seems that very small couplings would be sufficient to produce such an abundance and this
should make easier to have a metastable DM particle.  If one imposes that one of the seesaw neutrino in the model (\ref{Y+M})
is the DM particle,  then necessarily the neutrino Yukawa matrix must be in one of these three forms \cite{ad,unified}
\be\label{WASS}
 h^\nu \simeq
 \left( \begin{array}{ccc}
\ve_{e1} &  h_{ e 2}  &  h_{e 3}  \\
\ve_{\mu 1} &  h_{\m 2}  &  h_{\m 3} \\
\ve_{\tau 1} &  h_{\t 2}  &  h_{\t 3}
\end{array}\right) \,     , \,  {\rm or} \,
\left( \begin{array}{ccc}
h_{e 1} & \ve_{e 2}  &  h_{e 3}  \\
h_{\m 1} & \ve_{\mu 2}  &  h_{\mu 3} \\
h_{\t 1}  & \ve_{\tau 2} &  h_{\t 3}
\end{array}\right) \,   , \, {\rm or} \, 
\left( \begin{array}{ccc}
h_{e 1}    &  h_{e 2}  & \ve_{e 3}    \\
h_{\m 1} &  h_{\m 2} & \ve_{\mu 3} \\
h_{\t 1}  &  h_{\t 2} & \ve_{\tau 3}
\end{array}\right) \,   ,
\ee
where the Yukawa couplings $\ve_{\a I}$ of one RH neutrino, corresponding to the entries of one of the three columns, are tiny,
possibly because proportional to some small symmetry breaking parameters. We will refer to such very weakly coupled RH neutrino as
the {\em dark neutrino} and denote it by $N_{\rm D}$. The lifetime of the dark neutrino is in this case given by
\be\label{tauD}
\tau_{\rm D} = {4\pi \over h^2_A \, M_{\rm D}} \sim 10^{-26} \, h^2_A \, {\rm TeV \over M_{\rm D}} \, {\rm s} \,  , 
\ee
where $h^2_A = \sum_\a \ve_{\a {\rm D}}^2$ and the subscript ${\rm D}= 1,2$ or $3$, depending on the choice in (\ref{WASS}).  
Imposing the lower bound $\tau_{\rm D} \gtrsim \tau_{\rm D}^{\rm min} \sim 10^{28}\,{\rm s}$ from the IceCube neutrino telescope, as we will discuss, 
one finds  $h_A \lesssim 10^{-27}\,\sqrt{{\rm TeV}/M_{\rm D}}$. Such small values for the Yukawa couplings make impossible to 
find any efficient production mechanism for the dark neutrino.\footnote{Here I am assuming $M_{\rm D} \gg v \sim 100{\, {\rm GeV}}$.
On the other hand, we have seen that if $M_{\rm D} \ll m_e$ then the lifetime would be much suppressed compared to 
Eq.~(\ref{tauD}) and one finds the solution $M_{\rm D} \sim {\rm keV}$ with $N_{\rm D}$ produced by active-sterile neutrino mixing. 
In \cite{Datta:2021elq} the authors find that a MeV dark neutrino could satisfy the DM requirements,
being efficiently produced by gauge boson decays, though most recent 
constraints $X$-ray observations  seem to almost rule out this scenario.}

We can then conclude that with a minimal type-I seesaw extension of the SM, one cannot find a solution for a very heavy dark neutrino
playing the role of DM. We have then to consider a further extension. 

\section{Anisimov operators}

The idea is to consider an effective theory where the new physics does not need to be specified but it is encoded
in 5-dim (Anisimov) operators involving only RH neutrino and Higgs field \cite{anisimov,ad}. Therefore, one has now an 
effective Lagrangian given by
\be\label{effectiveL}
{\cal L}_{\rm eff} = {\cal L}_{\rm SM}  + {\cal L}^\nu_{Y+M} + {\cal L}_{\rm A} \,  ,
\ee
where
\be\label{anisimovop}
{\cal L}_A = \sum_{I, J}{\la_{I J} \over \L} \, \Phi^\dagger \, \Phi \, \overline{N_{\rm I}^c} \, N_{J}
= {\lambda_{\rm DS} \over \L} \, \Phi^\dagger \, \Phi \, \overline{N_{\rm D}^c} \, N_{{\rm S}} + 
                         {\lambda_{\rm SS} \over \L} \,\, \Phi^\dagger \, \Phi \, \overline{N_{\rm S}^c} \, N_{{\rm S}} +
                         {\lambda_{\rm DD} \over \L} \, \, \Phi^\dagger \, \Phi \, \overline{N_{\rm D}^c} \, N_{{\rm D}} \,  .
\ee
Here I am denoting by $N_{\rm S}$  one of the two seesaw neutrinos that reproduce the atmospheric and solar neutrino mass
scales. We can just consider the interaction with one of them, since the inclusion of the third one would just produce more stringent 
constraints. However this still has a cosmological role in producing the necessary interference with $N_{\rm S}$ 
to have non-vanishing $C\!P$ asymmetries. In this way the decays of the two seesaw neutrinos can produce a $B-L$ asymmetry that is partly converted
by sphalerons into the observed matter-antimatter asymmetry, so that one can also have successful leptogenesis.

The first term in ${\cal L}_A$ is the RH-RH (sterile-sterile) Higgs induced neutrino mixing (RHINO) operator
and will be responsible for the direct production of the dark neutrinos. Before discussing this production three interesting things 
can be noticed:
\begin{itemize}
\item The effective Lagrangian can be regarded as a step further compared to SMEFT \cite{talkdegrande}, where also RH neutrino fields
are included in the effective operators and it can be regarded as a specific example of $\nu$SMEFT \cite{talktalbert}.
\item The Anisimov operators are analogous to the Weinberg operator, a kind of further step in energy scale. 
\item They are a generalisation of the usual Higgs portal renormalisable operator involving a scalar \cite{Patt:2006fw}. 
\end{itemize}

\begin{figure}[t]
\centerline{\psfig{file=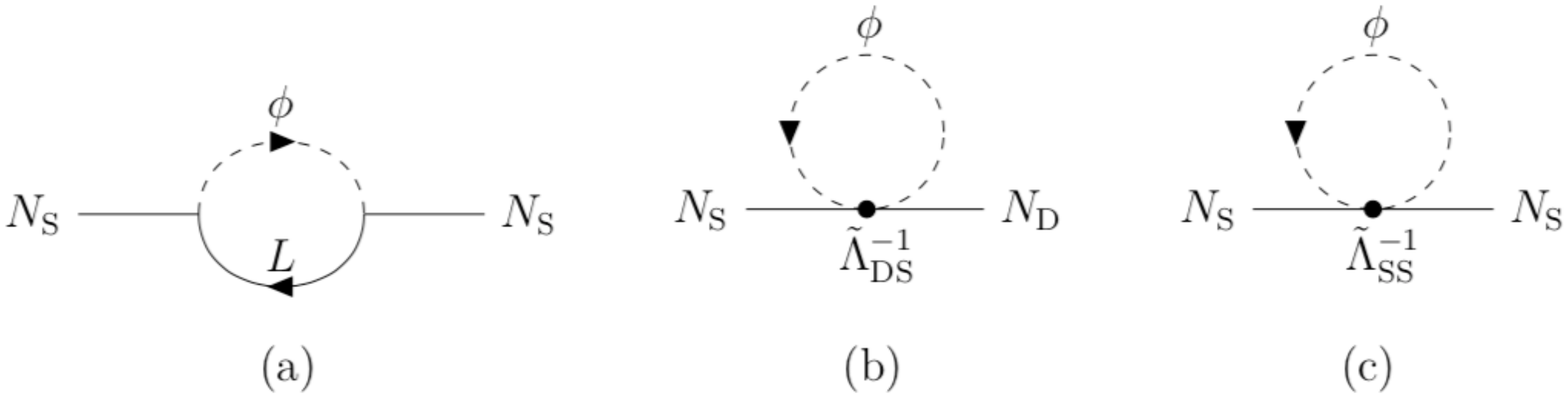,height=14cm,width=11cm,angle=-90}}\vspace{-5mm}
\caption{Self-energy diagrams from Yukawa interactions (panel (a)) and from Anisimov interactions (panels  (b) and (c)) \cite{DiBari:2022dtx}.}
\end{figure}

\section{RHINO dark matter}

I want now to focus on the RHINO operator, showing how its presence in the early universe can convert a small fraction of source neutrinos  
into dark neutrinos. One has then to check whether the dark neutrino relic abundance can match the observed DM abundance. First of 
all it is convenient to introduce the effective scale $\widetilde{\Lambda}_{\rm DS} \equiv \Lambda/\lambda_{\rm DS}$. If $\lambda_{\rm DS} \neq 0$,
then a source-dark (sterile-sterile) neutrino mixing is induced. 
One has to take into account finite temperature medium effects induced by the RHINO operator. 
These generate a contribution to the self-energy of the mixed source-dark neutrino system (see left panel in Fig.~1). 
Also source neutrino Yukawa self-interactions contribute to the self energy and need to be taken into account (see right panel in Fig.~1). These self-energy 
contributions can be also conveniently described  in terms of effective potentials $V_{\rm DS}^\Lambda$ and $V_{\rm SS}^Y$, respectively, given by
\be
V^{\L}_{\rm DS} = \frac{T^2}{12\,\widetilde{\Lambda}_{\rm DS}} \;\;\; \mbox{\rm and} \;\;\;
V^{Y}_{\rm SS} =  \frac{T^2}{8\,E_J} \, h^2_{\rm S} \,  .
\ee
These need to be taken into account in the effective mixing Hamiltonian describing the evolution of the mixed system given by
\be\label{effectiveham}
\Delta {\cal H} \simeq   
\left( \begin{array}{cc}
- \frac{\D M^2}{4 \, p} - \frac{T^2}{16\,p} \, h^2_{\rm S} &  \frac{T^2}{12\,\widetilde{\L}_{\rm DS}}  \\[1ex]
\frac{T^2}{12\,\widetilde{\L}_{\rm DS}} &  \frac{\D M^2}{4 \, p} + \frac{T^2}{16 \, p} \, h^2_{\rm S}  
\end{array}\right)  \, ,
\ee
where we used the ultrarelativistic approximation\footnote{This assumes that the dark neutrino production
occurs in the ultra-relativistic regime, as it will be verified.} and 
defined $\D M^2 \equiv M^2_{\rm S} - M^2_{\rm D}$. 

The production of dark neutrinos can be described by the density matrix equation \cite{densitym}
\be\label{densitymatrixeq}
{d {{\cal N}} \over dz} = -{i\over H(z)z}\,[\D{\cal H}, {\cal N}]  - 
\begin{pmatrix}
0   &  {1\over 2}(D+S) \,{\cal N}_{\rm DS}  \\ 
{1\over 2}(D+S)  \,{\cal N}_{\rm SD}  & (D+S)\,(N_{N_{\rm S}}-N_{N_{\rm S}}^{\rm eq})  
\end{pmatrix} \,   ,
\ee
where we defined $z \equiv M_{D}/T$ normalised the density matrix ${\cal N}$ in a way that the diagonal elements 
give the abundances of dark neutrinos, $N_{N_{\rm D}} = {\cal N}_{\rm DD}$, and
source neutrinos, $N_{N_{\rm S}} = {\cal N}_{\rm SS}$. Notice how decays and scatterings also contribute
to decoherence effects, damping the density matrix off-diagonal terms.
Initially, for $z =z_{\rm in}$, the density matrix is simply given by:
\be
{\cal N}(z_{\rm in}) = N_{N_{\rm S}}(z_{\rm in})\,\left(\begin{array}{cc}
0 & 0 \\
0 & 1
\end{array}\right) \,  .
\ee 
Here notice that the abundances are normalised in a way that in ultrarelativistic thermal equilibrium one has $N_{N_i}^{\rm eq}(T \gg M_i) = 1$
and consequently $N_{\gamma} = 4/3$. Notice that with this normalisation the 
final value of dark neutrino abundance that is necessary to reproduce the
measured value of DM energy density parameter is given by
\be\label{NDobs}
N_{N_{\rm D}}^{\rm f,obs} \simeq 1.1 \times 10^{-7} \, {{\rm GeV} \over M_{\rm D}} \,  .
\ee
\begin{figure}
\begin{center}
\psfig{file=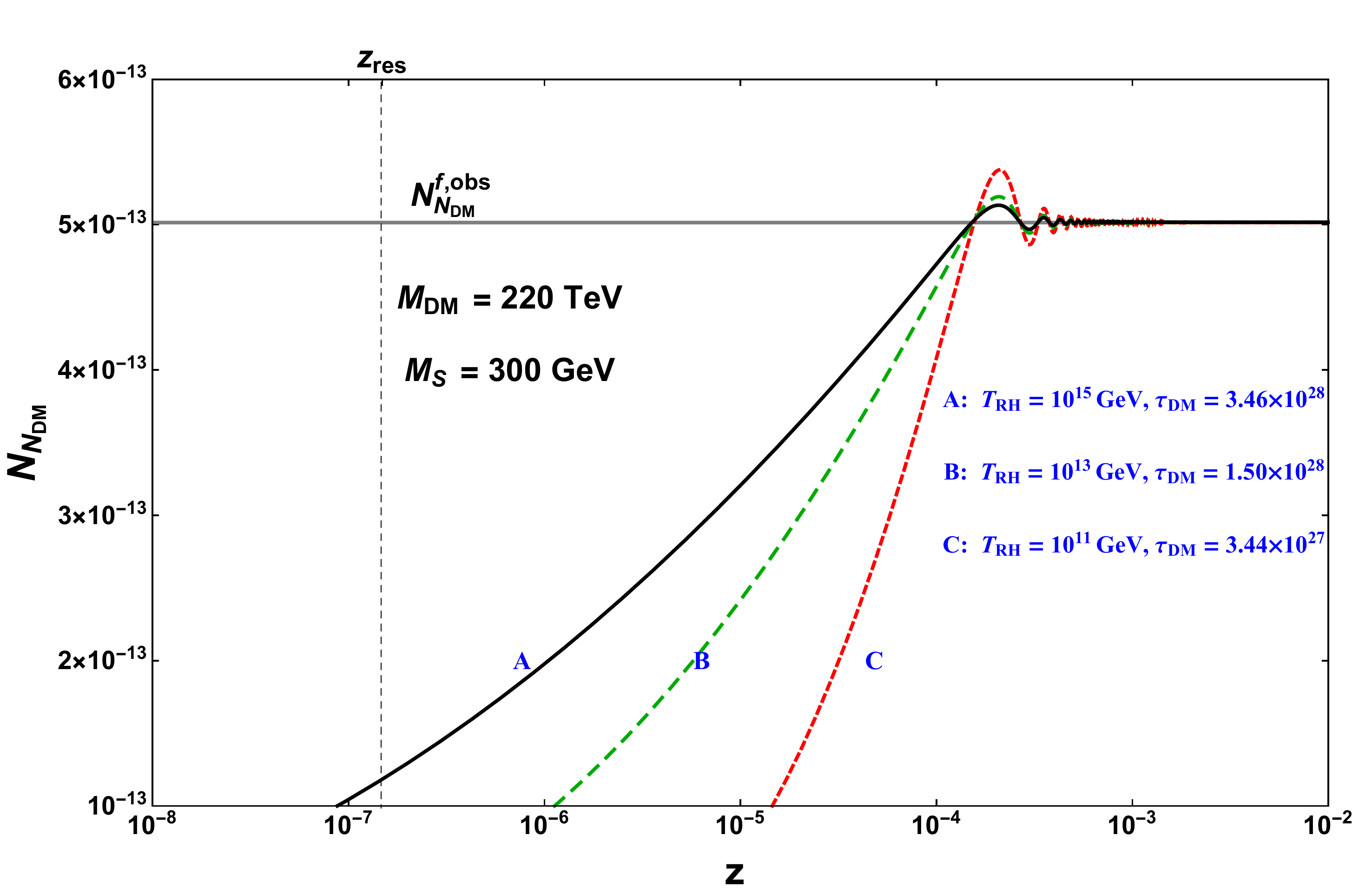,height=85mm,width=125mm} 
\end{center}
\vspace{-1mm}
\caption{Evolution of the DM abundance $N_{N_{\rm DM}}$ for three different choices
of $T_{\rm RH}$ and $\tau_{\rm DM}$ as indicated in a linear plot for the abundance (from \cite{densitym}).}
\label{benchmark}
\end{figure}
In Fig.~2 we show the production of the dark neutrino abundance for fixed values $M_{\rm D} = 220\,{\rm TeV}$ and $M_{\rm S} = 300 \, {\rm GeV}$
and for three choices of the values of the parameters $T_{\rm RH}$ and $\tau_{\rm D}$ as indicated, 
such that the relic abundance reproduces the value $N_{N_{\rm D}}^{\rm f,obs} \simeq 5 \times 10^{-13}$.
Moreover, as we will see, the values of the lifetime $\tau_{\rm D}$ respect the lower bound from IceCube measurement of the high energy neutrino flux.
There are two important things to notice in these solutions:
\begin{itemize}
\item An {\em initial thermal abundance of source neutrinos}, i.e., $N_{N_{\rm S}}^{\rm in} = 1$, is assumed. I will show soon how this assumption can be justified within a logical extension of the scenario.
\item The dark neutrino is assumed to be heavier than the source neutrino, i.e., $M_{\rm D} > M_{\rm S}$. All results I present in this talk
are obtained for such a choice. The case $M_{\rm D} < M_{\rm S}$ requires a dedicated analysis that will be presented elsewhere.
\end{itemize}

\section{Constraints from decays}

The RHINO operator is also responsible for the instability of the dark neutrinos that have necessarily to decay \cite{ad,unified}. 
With the assumption of heavy RH neutrinos, $M_I \gtrsim 100\,{\rm GeV}$, one has two dominant decay channels. 
The first is {\em two body decay} channel where the dark neutrino, via the mixing with the source neutrino, decays into a gauge boson 
$A=Z_0, W^{\pm}, \gamma, H$ and a charged lepton or neutrino with decay rate  $\G_{{\rm D} \ra A + \ell_{\rm S}}$.
The second channel is {\em four body decay}, such that $N_{\rm D} \ra 3 A + N_{\rm S}$,  with decay rate $\G_{{\rm D} \ra 3A + {\ell}_{\rm S}}$.
The diagrams for these two channels are shown in the left and right panel, respectively, of Fig.~3.  Three body decays are also possible but give a sub-dominant channel that can be neglected.
\begin{figure}
\begin{center}
\psfig{file=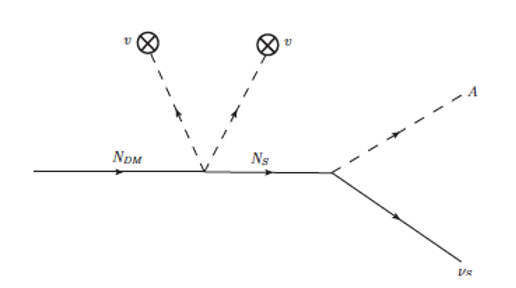,height=45mm,width=60mm}
\psfig{file=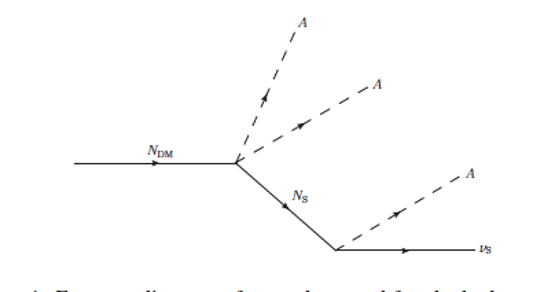,height=45mm,width=60mm}
\end{center}
\vspace{-1mm}
\caption{Diagrams for two (left) and four (right) body decay (from \cite{densitym}).}
\label{benchmark}
\end{figure}

In the case of the two body decay,  the dark neutrino mixes today, at zero temperature, with the source neutrino with a mixing angle 
\be\label{thetaL0}
\theta_{\L 0} = \frac{2\, v^2/\widetilde{\L}_{\rm DS}}{M_{\rm D}\,(1-M_{\rm S}/M_{\rm D})} \,  .
\ee
In this way the two body decay rate is given by
\be\label{twobody}
\Gamma_{{\rm D} \ra A + \ell_{\rm S}}  = 
{h^2_{\rm S} \over \pi} \, \left(\frac{v^2}{\widetilde{\L}}\right)^2
\, {M_{\rm D}\over (M_{\rm D} - M_{\rm S})^2}\,  .
\ee
The four body decay rate can be calculated in the narrow width approximation, obtaining
\be
\G_{{\rm D} \ra 3A + {\ell}_{\rm S}} = 
{\Gamma_{\rm S} \over 15 \cdot 2^{11} \cdot \pi^{4}} \, {M_{\rm D} \over M_{\rm S}} \, 
\left({M_{\rm D} \over \widetilde{\L}_{\rm DS}}\right)^2 \,  ,
\ee
where $\G_{\rm S} = h^2_{\rm S}\,M_{\rm S}/(4\,\pi)$.  The lifetime can then be calculated as
\be\label{lifetimebound}
\tau_{\rm D} \simeq (\G_{{\rm D} \ra A + \ell_{\rm S}}  + \G_{{\rm D} \ra 3A + {\ell}_{\rm S}})^{-1} \,  .
\ee
As I am going to discuss in more detail, IceCube data impose roughly a lower bound on $\tau_{\rm D}$.
As one can see from the expressions of the rates, this implies that two body decays place a lower bound 
on $M_{\rm D}$ while four body decay place an upper bound. In this way there is an intrinsic  allowed range 
for the dark neutrino mass that emerges from the model when the lower bound is imposed.
\begin{figure}
\begin{center}
\psfig{file=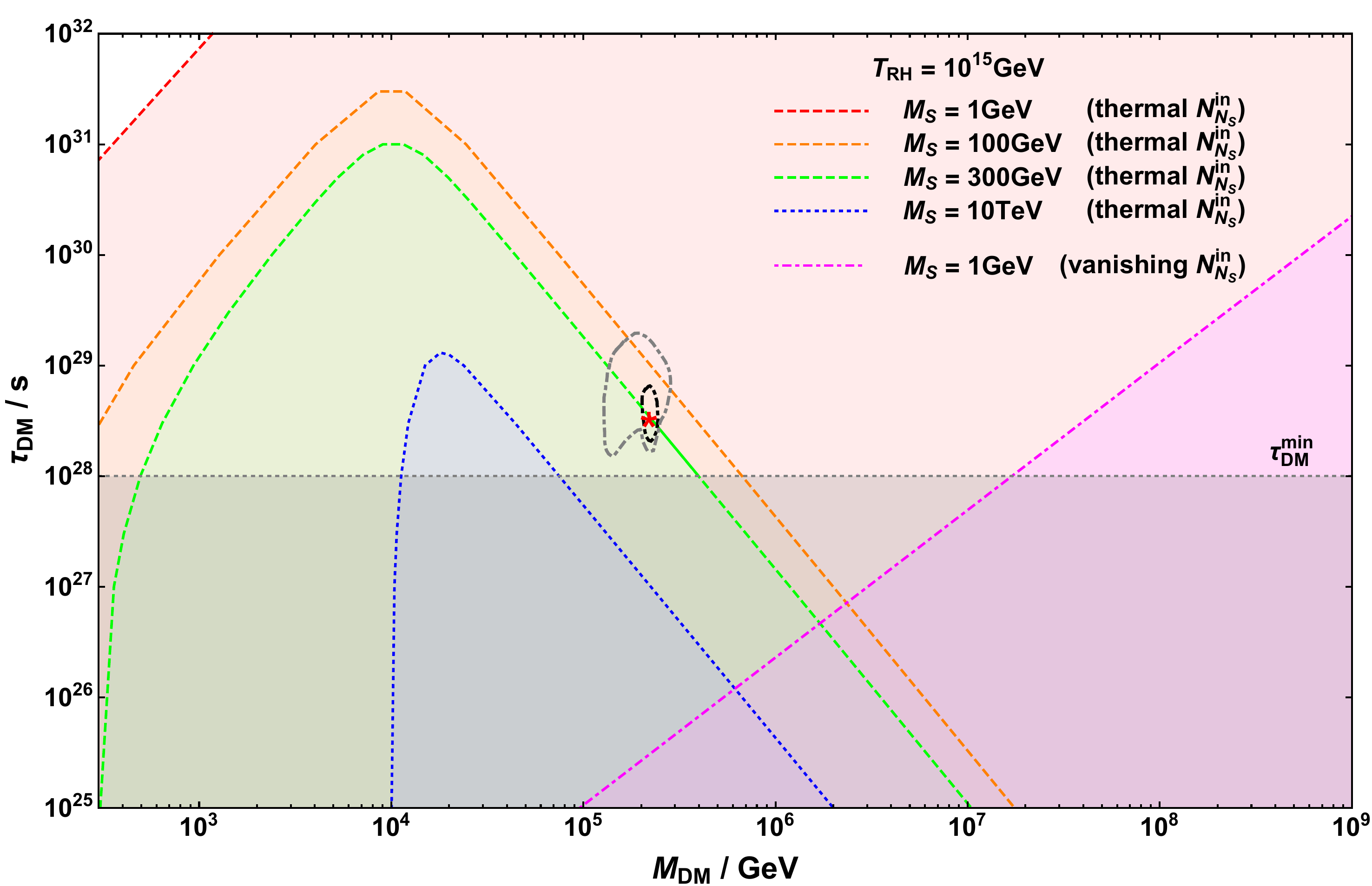,height=65mm,width=125mm} 
\end{center}
\vspace{-1mm}
\caption{Allowed regions in the plane $\tau_{\rm DM}$ versus $M_{\rm DM}$ for different values of $M_{\rm S}$ 
and for $T_{\rm RH} = 10^{15}\,{\rm GeV}$ (from \cite{densitym}).}
\label{benchmark}
\end{figure}
In Fig.~4 we show the allowed regions in the plane $\tau_{\rm D}$ vs. $M_{\rm D}$. Notice that there are four regions obtained
for initial thermal $N_{\rm S}$ abundance. They correspond to four different values of $M_{\rm S}$ as indicated. One can see that
for an increasing value of $M_{\rm S}$, since the lifetime decreases, as it can be easily inferred from Eqs.~(\ref{twobody})-(\ref{lifetimebound}),
the allowed region shrinks.  I will comment soon on the importance of the value of $M_{\rm S}$ for leptogenesis. First,
however, at this point I notice that one could legitimately have an objection: the allowed regions have been obtained 
assuming arbitrarily that the source neutrino have a thermal abundance at the time of the mixing with the dark neutrinos. 
If we relax this assumption and consider an initially vanishing source neutrino abundance, then one can see that Yukawa interactions
are by far insufficient to produce a source neutrino abundance close to  a thermal value prior to the mixing. The consequence is quite
dramatic: all the shown allowed regions simply evaporate. The only possibility is then to consider a value $M_{\rm S} < M_{\rm W} \simeq 80 \,{\rm GeV}$
in a way that the decay rate of source neutrinos decreases significantly and an allowed region appears. In Fig.~4 the case $M_{\rm S} =1\,{\rm GeV}$
is shown. One can see that the allowed region now appears for $M_{\rm D} \gtrsim 10\,{\rm PeV}$, so for much heavier dark neutrinos. 
This is because the upper bound on $M_{\rm D}$ from four body decays essentially disappears and one is left only with a much more stringent lower bound from two body decays. The situation  is clearly much less attractive  now and so one can have a legitimate question: is there any process
that can thermalise the source neutrino abundance prior to the mixing with the dark neutrinos thus opening the (much more attractive) allowed
regions at much higher values of $M_{\rm S}$ and lower values of $M_{\rm D}$?

There are two very good motivations to try to investigate this possibility. I am going first to discuss these two motivations 
and then as a last point of my talk  I will finally show that indeed there is a solution to such an issue and it is actually quite a
natural one, not artificially imposed.  

\section{Unified picture for the origin of matter in the universe}

The value of $M_{\rm S}$ is particularly interesting if one wants to combine leptogenesis, solving the problem of the origin of matter within the same model, 
since it sets the scale of leptogenesis. It can be seen how for value above $10\,{\rm TeV}$ the allowed regions in the case of initially 
thermal $N_{\rm S}$ abundance tend to disappear. The minimum value of $M_{\rm S}$ for (resonant) leptogenesis to work in a way 
to be independent of the initial conditions is approximately $M_{\rm S}\sim 300\,{\rm GeV}$, since below this value, the $B-L$ asymmetry is generated 
in the non-relativistic regime, at the end of the wash-out regime, too late to be converted into a baryon asymmetry. One has in this case to resort to contributions produced in the ultra-relativistic regime that strongly depend on the initial conditions. Therefore, there is a wide
allowed range $M_{\rm S} \sim 300\,{\rm GeV}$--$100\,{\rm TeV}$ that allows a solution to the problem of the origin of matter of the universe
if one can justify an initial thermal $N_{\rm S}$ abundance. In Fig.~5 one can see a particular realisation of a simultaneous generation of
dark neutrino abundance and $B-L$ asymmetry that reproduce the measured values of DM energy density parameter and baryon-to-photon number ratio.  The possibility to realise such a  picture is particularly intriguing but one needs to find a way to thermalise the source neutrinos prior to the mixing. 
Let us first discuss a second equally intriguing motivation that is related to the possibility to test the RHINO model. 
\begin{figure}
\begin{center}
\psfig{file=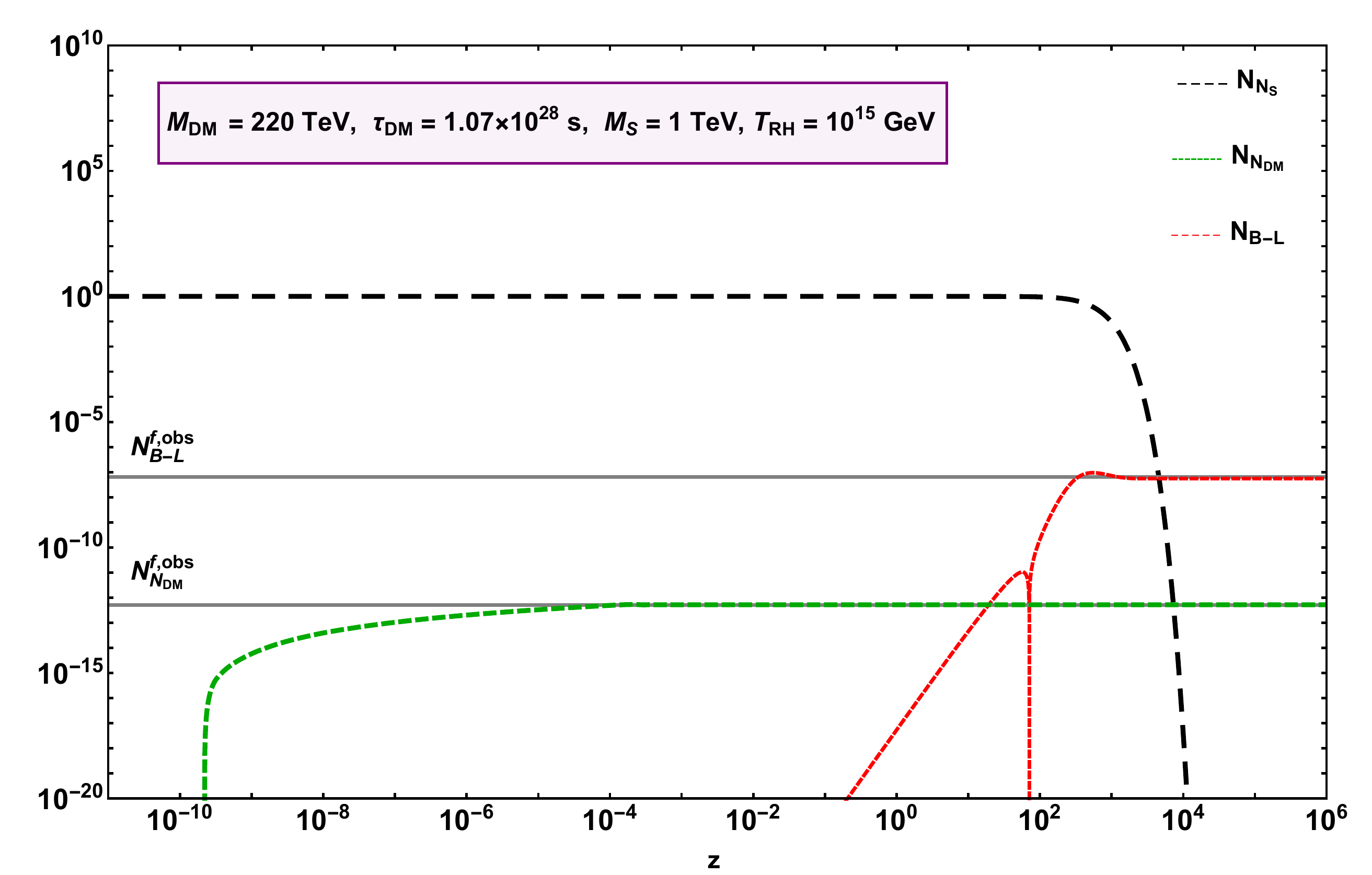,height=85mm,width=115mm} 
\end{center}
\vspace{-1mm}
\caption{Evolution of the $B-L$ asymmetry and 
DM abundance $N_{N_{\rm DM}}$ for a choice of parameters such that the final values simultaneously
reproduce the observed values of matter-antimatter asymmetry and DM abundance. 
The $N_{\rm S}$-abundance is also shown. One can see that the plot 
is for initial thermal $N_{\rm S}$-abundance (from \cite{densitym}).}
\label{benchmark}
\end{figure}

\section{IceCube data and the 100 TeV anomaly}

Neutrinos are perfect astronomical messengers since they can travel to us from the edge of the universe \cite{Halzen:2002pg}.
In the energy range 10 TeV--10 EeV they are the only particles that can travel unabsorbed and undeflected. In 2013
IceCube discovered very high energy neutrinos (30 TeV -- 1 PeV range) \cite{icecube}. Some have been observed in coincidence with blazar $\gamma$-ray flare, an evidence of the presence of a component with extragalactic origin. IceCube employs two strategies to reduce the overwhelming 
atmospheric background at energies $\lesssim 300\,{\rm TeV}$. A first strategy is to impose a High Energy Starting Events (HESE) veto. 
This sample is the first that found evidence of a diffuse extraterrestrial neutrino component. A second strategy, more traditional, is based on
analysing an up-going muon data set. This has now also confirmed the existence of a diffuse  extraterrestrial neutrino component. However, there is a
a tension between the two data sets. The up-going muon sample energy spectrum is well described by a single power law $\propto E^{-\gamma_{\rm astro}}$
with spectral index $\gamma_{\rm astro} = -2.19$ \cite{Haack:2017dxi}. 
This is quite a standard astrophysical component well explained by a Fermi model for the acceleration of cosmic rays.  On the other hand, the case of HESE data, one would need a two-component power law spectrum, one {\em hard} component with a value of the spectral index compatible with the one from up-going muon sample dominating at energies above $\sim 300 \,{\rm TeV}$ and a second  {\em soft} 
component with spectral index $\gamma_{\rm soft} \simeq 3.7$ dominating at energies $E \sim 100\,{\rm TeV}$ \cite{IceCube:2017zho}. This second component is more difficult to understand in astrophysical models but not particularly surprising. What makes it challenging to be understood in terms
of an astrophysical solution is that such a soft component would be incompatible within a multimessenger analysis showing that the $\gamma$-ray flux at   $\sim 100\,{\rm GeV}$ energies (as measured by the Fermi satellite) would be comparable with the hard component flux as expected within traditional astrophysical models but it would be incompatible with the soft component that seems to suggest the existence of {\em hidden} sources producing just neutrinos and not photons. Similar conclusion can be also drawn analysing ultra-high energy cosmic rays spectrum. Unknown and quite exotic astrophysical hidden sources might be a solution to this anomaly. 
It is also intriguing that such unexplained excess can be well reproduced within the RHINO model \cite{unified}.   In Fig.~6 one can see an example how the decays of dark neutrino with mass $M_{\rm D}=300\,{\rm TeV}$ can reproduce the excess and help fitting IceCube data (in this 
2016 analysis, $\sim$4 year HESE data were employed). 
\begin{figure}
\begin{center}
\psfig{file=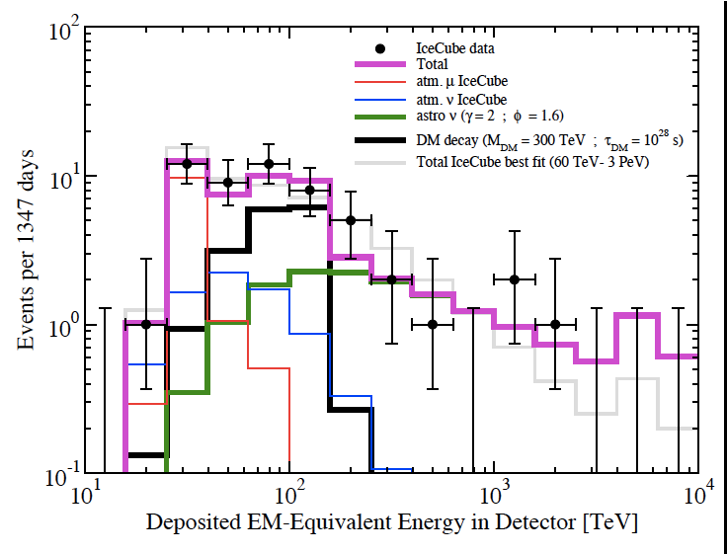,height=85mm,width=115mm} 
\end{center}
\vspace{-1mm}
\caption{Event spectra in the IceCube detector after 1347 days including a contribution from RHINO DM decays
with $M_{\rm D} = 300\,{\rm TeV}$ (from \cite{unified}).}
\label{benchmark}
\end{figure}
Recently, the IceCube collaboration has presented an analysis where IceCube 7.5 year data are fitted including a contribution 
to the neutrino flux from decaying DM into various channels, finding that with such an addition the  fit improves
at $2.5 \s$  significance level compared to the null hypothesis, corresponding to the case where 
just a power law astrophysical component is present \cite{IceCube:2022clp}. A decaying DM scenario it is then
a viable way to explain this excess at 100 TeV. Moreover, collection of more data should allow to perform tests
on the level of the anisotropies in the flux of very high energy neutrinos.  This would allow to disentangle
astrophysical explanations in terms of exotic hidden sources from a DM scenario, since anisotropies should show the presence
of a component tracking the DM distribution. 

In conclusion, the possibility to combine leptogenesis with RHINO DM in a unified model of the origin of matter of the universe and 
the possibility to provide an explanation for the 100 TeV excess in very high energy neutrino data, provide two strong motivations to
answer the question whether it is possible to find processes that can thermalise the source neutrino abundance prior to the
dark neutrino production from their mixing. 

\section{Including Higgs portal interactions for the source neutrino: a RHINO miracle}

Let us go back to the full set of Anisimov operators in Eq.~(\ref{anisimovop}). We have so far neglected
Higgs portal interactions producing the source neutrinos. These are natural processes to be included 
that might lead to a thermalisation of the source neutrinos prior to the mixing with the dark neutrinos \cite{DiBari:2022dtx}.
Therefore, we now consider the effective Lagrangian:
\be\label{effectiveL}
{\cal L}_{\rm eff} = {\cal L}_{\rm SM}  + {\cal L}^\nu_{Y+M} + 
{1 \over \widetilde{\Lambda}_{\rm DS}} \, \Phi^\dagger \, \Phi \, \overline{N_{\rm D}^c} \, N_{{\rm S}} + 
                         {1\over \widetilde{\Lambda}_{\rm SS}} \,\, \Phi^\dagger \, \Phi \, \overline{N_{\rm S}^c} \, N_{{\rm S}} \,  ,
\ee
where $\widetilde{\Lambda}_{\rm SS} \equiv \lambda_{\rm SS} / \Lambda$. The density matrix equation (\ref{densitymatrixeq}) now needs to be 
modified to include these additional interactions, becoming:
\be\label{densitymatrixeq}
{d {{\cal N}} \over dz} = -{i\over H(z)z}\,[\D{\cal H}, {\cal N}]  - 
\begin{pmatrix}
0   &  {1\over 2}(D+S) \,{\cal N}_{\rm DS}  \\ 
{1\over 2}(D+S)  \,{\cal N}_{\rm SD}  & (D+S)\,(N_{N_{\rm S}}-N_{N_{\rm S}}^{\rm eq})  + A \, (N^2_{N_{\rm S}} - N^{{\rm eq}\,2}_{N_{\rm S}}) 
\end{pmatrix} \,   ,
\ee
where 
\be
A \equiv {\langle \sigma_{\phi\phi^\dagger \ra N_{\rm S}N_{\rm S}^c}\, v_{\rm rel} \rangle \over H(z)\,z \, R^3(z)} \,  ,
\ee
and $\langle \sigma_{\phi\phi^\dagger \ra N_{\rm S}N_{\rm S}^c}\, v_{\rm rel} \rangle$ is the thermal averaged cross section.
Here $R^3(z)$ is the portion of comoving volume where abundances are calculated, essentially a normalisation factor. 

The thermal averaged cross section is simply given by \cite{Kolb:2017jvz}
\be
\left.\langle \sigma_{\phi\phi^\dagger \ra N_{\rm S}N_{\rm S}}\,v_{\rm rel} \rangle\right|_{M_{\rm S}\ll T} 
= {1 \over 4 \pi \, \widetilde{\Lambda}_{\rm SS}^{2} } \,  .
\ee
One can rewrite $A(z)$ as
\be
A(z) = {A_1 \over z^2} \,  , \;\; \mbox{\rm with} \;\;
A_1 \equiv A(z=1) = {3 \over 16}\, {\zeta(3) \over \pi^3}\,\,g_{N_{\rm S}}\,\sqrt{90\over 8\,\pi^3\,g_R}\, 
{M_{\rm D}\, M_{\rm P}\over \widetilde{\Lambda}_{\rm SS}^{2}} 
\ee
and a convenient numerical expression for $A_1$ is given by
\be\label{A1num}
A_1 \simeq 1.0 \times 10^{-11}\,\left({M_{\rm D}\over 100\,{\rm TeV}}\right)\,
\left({10^{16}\,{\rm GeV}\over \widetilde{\Lambda}_{\rm SS}}\right)^2 \,  .
\ee
From this expression for $A(z)$ one can then finally calculate the source neutrino abundance prior to the mixing given by
\be\label{NSabundance}
N_{N_{\rm S}}(z_{\rm in} \ll z \ll 1) - N_{N_{\rm S}}(z_{\rm in}) \simeq {A_1\over z_{\rm in}} 
\simeq 1.0 \times \left({T_{\rm in}\over 10^{16}\,{\rm GeV}}\right) 
\,  \left({10^{16}\,{\rm GeV} \over \widetilde{\Lambda}_{\rm SS}}\right)^2 \,  .
\ee
This numerical expression already highlights the emergence of the grand-unified scale as the natural scale
of new physics for the effectiveness of these interactions in thermalising the source neutrinos. 

In Fig.~7 we show a benchmark case of evolution of the abundances of source and dark neutrinos including both the
RHINO operator and Higgs portal interactions for the source neutrinos. In this example we have
taken $\widetilde{\Lambda}_{\rm SS} = 10^{16} \, {\rm GeV}$ and $\widetilde{\Lambda}_{\rm DS} = 10^{23}\,{\rm GeV}$.
These are typical values leading to successful dark neutrino production and stability.
\begin{figure}[t]
\centerline{\psfig{file=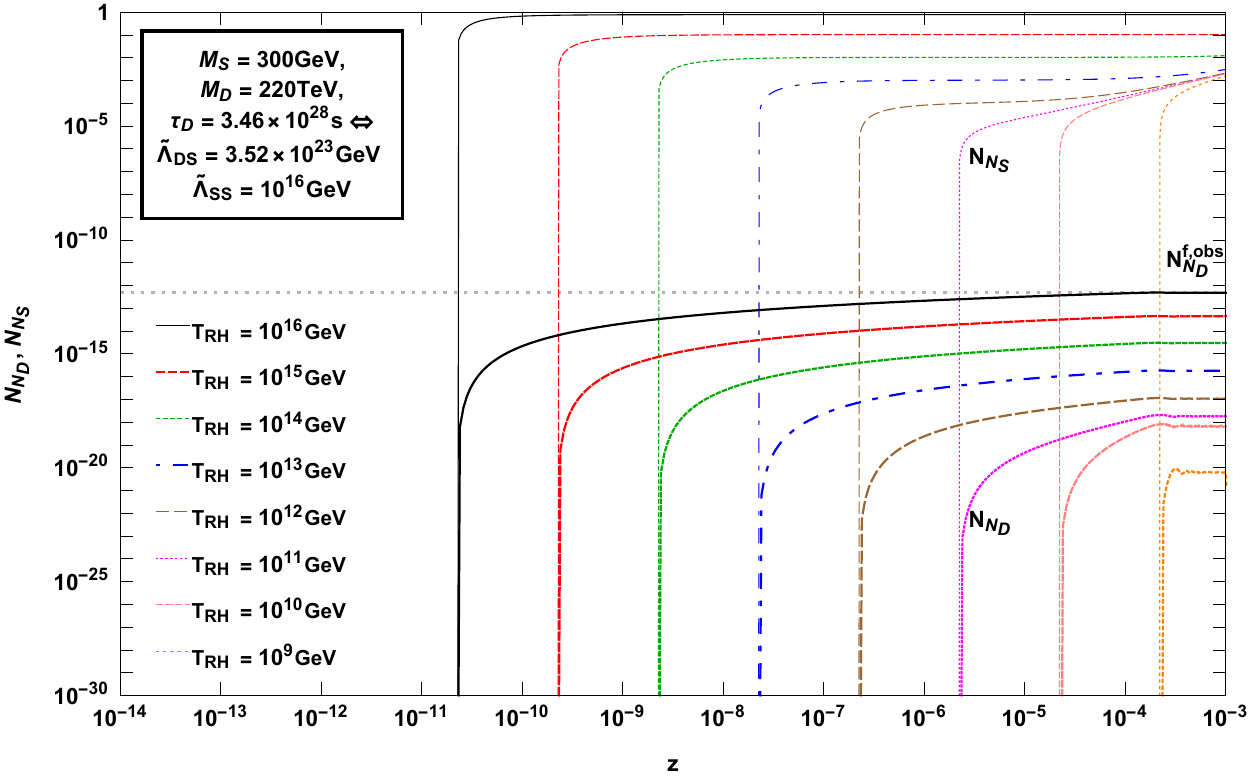,height=6cm,width=9cm,angle=0}}
\caption{Effect of source neutrino Higgs portal interactions on the source and dark neutrino abundances (from \cite{DiBari:2022dtx}).}
\end{figure}

In Fig.~8 the allowed regions in the traditional plane $\tau_{\rm D}$ versus $M_{\rm D}$ are shown as in Fig.~5. This time
there is no assumption of initial thermal source neutrino abundance, the job is done by the Higgs portal interactions. 
In this figure the effective scales  $\widetilde{\Lambda}_{\rm SS} = 10^{16} \, {\rm GeV}$ and $M_{\rm S} =300\,{\rm GeV}$.
The regions are shown for different value of $T_{\rm RH}$. In Fig.~9 the same regions are shown for $M_{\rm S} = 10\,{\rm TeV}$
and $M_{\rm S} = 100\,{\rm TeV}$ and again one can see that the latter gives the maximum value of the seesaw scale corresponding to
the scale of leptogenesis. At these scales leptogenesis needs to be resonant \cite{resonant} but the final asymmetry can be independent of the
initial conditions.  In the figures one can also see lower bounds on $\tau_{\rm D}$ placed by different data sets and assuming different decay 
channels for DM.  The star denotes the best fit found in \cite{IceCube:2022clp} for 7.5 year IceCube HESE data. The model can easily reproduce
the best fit value for $M_{\rm D} = 386 \, {\rm GeV}$. 
\begin{figure}[t]
\centerline{\psfig{file=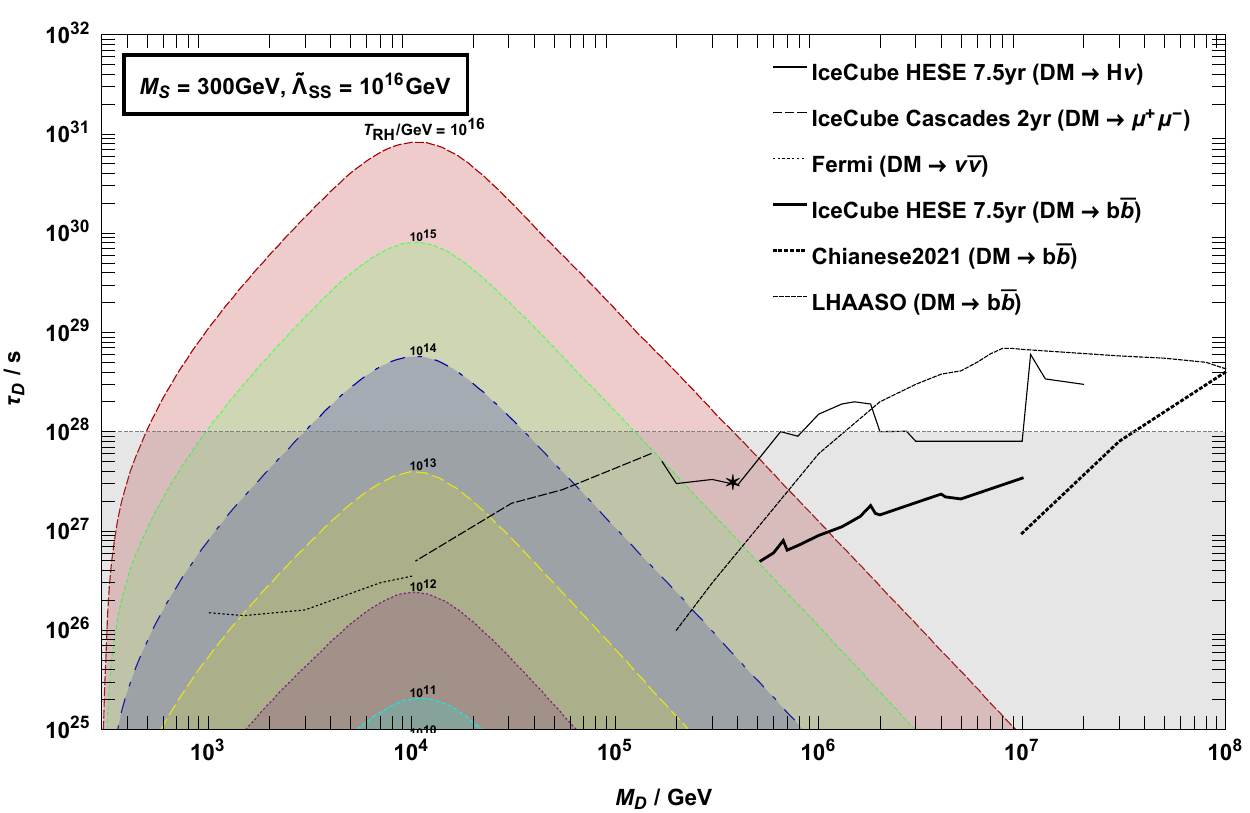,height=6cm,width=10cm,angle=0}}
\caption{Allowed regions in the lifetime versus mass plane for a fixed value $\tilde{\Lambda}_{\rm SS} = 10^{16}\,{\rm GeV}$, 
for the indicated values of $T_{\rm RH}$ and for $M_{\rm S}=300\,{\rm GeV}$ (upper panel), $1\,{\rm TeV}$ (bottom panel) \cite{DiBari:2022dtx}.}
\end{figure}
The choice of the grandunified scale for $\widetilde{\Lambda}_{\rm SS}$ maximises the allowed region in the parameter space
and emerges as the favourite scale of new physics in the RHINO model and this can legitimately be seen as a very encouraging 
coincidence, what I refer to as the {\em RHINO miracle}, paraphrasing the old WIMP miracle. 
However, there is another important objection and potentially unsatisfactory feature. In Fig.~7 we have seen that the typical necessary 
values of the 
effective scales $\widetilde{\Lambda}_{\rm SS} \sim 10^{16}\,{\rm GeV}$ and 
$\widetilde{\Lambda}_{\rm DS} \sim 10^{23}\,{\rm GeV}$ differ by many orders of magnitudes.
It would seem that a successful production of dark neutrinos able to respect all DM experimental requirements necessarily requires
two different new physics scales. This complications, and increase of parameters of the model, 
looks as a price to pay for it to work, providing a successful picture of the origin of matter of the universe.  
However, as I am going to discuss, there is  a simple UV-completion clearly showing it is possible to have 
just one scale of new physics and strikingly this coincides with the grandunified scale.  

\section{UV-completions: a possible GUT origin}

I will now discuss two possible UV-completions for the RHINO models including Higgs portal interactions
for the source neutrinos. In a first case the mediator of the interactions is a very heavy Higgs scalar \cite{ad,unified,Kolb:2017jvz,DiBari:2022dtx} 
and in the second case a very heavy fermion \cite{ad,DiBari:2022dtx}. As we will see, while the first option leads to quite a rather contrived choice of parameters, the second option is strikingly simple and successful in addressing all issues. 

\subsection{Heavy scalar as a mediator}

In a first extension of the seesaw Lagrangian one introduces a heavy real scalar field $H$ (with vanishing vev) coupling to
the RH neutrinos with Yukawa couplings $y_{IJ}$ and to the standard Higgs field with a trilinear coupling $\mu$:
\be
{\cal L}_H = {1\over 2}\partial_\mu H \partial^\mu H -{1\over 2}\,M^2_H \, H^2 - \lambda_{IJ}\,H \, \overline{N_{\rm I}^c} \, N_{J} 
- \mu \, H \, \phi^\dagger \, \phi \,  .
\ee
At scales much below $M_H$ we can integrate out $H$, obtaining the effective Lagrangian
\be
{\cal L}_H^{\rm eff}= 
{1\over 2}\,\sum_{I,J,K,L} {\lambda_{IJ}\lambda_{KL}\over M^2_H} \, (\overline{N_{\rm I}^c} \, N_{J})\,(\overline{N_{\rm K}^c} \, N_{L}) 
+ {1\over 2}\,{\mu^2 \over M^2_H}\,(\phi^\dagger \, \phi)^2
+ {\mu\,\lambda_{IJ} \over M^2_H}\, \, \Phi^\dagger \, \Phi \, \overline{N_{\rm I}^c} \, N_{\rm J} \,  .
\ee
One can clearly recognise the Anisimov operators in Eq.~(\ref{anisimovop}) and the effective scales can 
be identified with $\widetilde{\Lambda}_{IJ} = \Lambda/\lambda_{IJ}$,
and $\Lambda = M^2_H/\mu$.  Diagrammatically, the self-energy diagram in the panel (b) of  Fig.~1
and a four-point interaction scattering diagram (not shown) are obtained by the diagrams in Fig.~9, panel (a) and panel (b),  respectively. 
\begin{figure}[t]
\centerline{\psfig{file=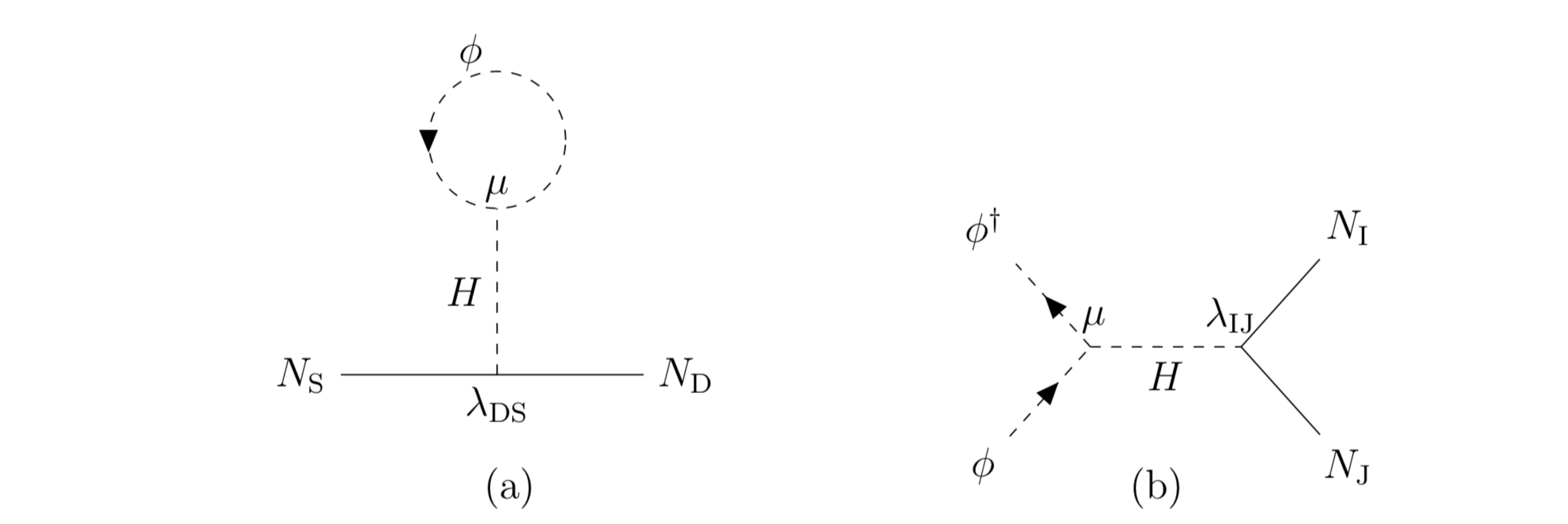,height=6.5cm,width=19cm,angle=0}}\vspace{-2mm}
\caption{Feynman diagrams with a heavy scalar $H$ as mediator and $I,J={\rm D,S}$. Integrating out $H$, they 
lead to the Feynman diagrams in panel (b) of Fig.~2 and  (a), (b) and (c) of Fig.~3.}
\end{figure}
The appealing feature of this model is that one can get a trans-Planckian value for the effective scale $\widetilde{\L}_{\rm DS} \sim 10^{23}\,{\rm GeV}$ 
even for $\lambda_{IJ} = {\cal O}(1)$, simply choosing $\mu \ll M_{\rm GUT}$. For example, one can take 
$M_H \sim M_{\rm GUT} \sim 10^{16}\,{\rm GeV}$
and $\mu \sim 10^9\,{\rm GeV}$. However, the problem of this setup is that one cannot also reproduce the effective scale 
$\widetilde{\Lambda}_{\rm SS} \sim 10^{16}\,{\rm GeV}$ for the source neutrino Higgs portal interactions. 
In  that respect, one should  arbitrarily assume 
$\Lambda \sim 10^{16}\,{\rm GeV}$, for example for $\mu = M_H \sim M_{\rm GUT} \sim 10^{16}\,{\rm GeV}$, 
 $\lambda_{\rm DS} \sim 10^{-7}$ and $\lambda_{\rm SS}\ll 10^{-7}$ in order for  
$\widetilde{\Lambda}_{\rm DD}$ to be sufficiently large that Higgs portal interactions producing dark neutrinos can be neglected (as we did).
This is setup is quite contrived and not really justified by any plausible argument.
 
There is a much simpler model where the values of the needed values of the effective scales emerge quite naturally.

\subsection{Heavy fermion $F$ as mediator}

 Let us extend the seesaw Lagrangian introducing an heavy fermion doublet $F$ with Yukawa couplings $y_I$ to RH neutrinos, 
 \be
 {\cal L}_F = \bar{F}\, ( i\, \slashed{\partial} - M_{\rm F} ) \, F - y_{I}\, (\bar{F} \, \phi \, N_{I} + \bar{N}_I\,\phi^\dagger \, F) \,  .
 \ee
At scales much below $M_{\rm F}$ one can integrate out  $F$ obtaining the effective Lagrangian
\be
-{\cal L}_{F}^{\rm eff} = {y_I \, y_J \over M_F} \, \bar{N}_I\,N_J\,\phi^\dagger\,\phi \,  ,
\ee
where the RH side coincides with the Anisimov operators with the simple identification $\L = M_{\rm F}$
and $\lambda'_{IJ} = y_I\,y_J$. The three Anisimov operators in Eq.~(\ref{anisimovop}), Higgs-induced neutrino mixing, source neutrino
Higgs portal interactions and dark neutrino Higgs portal interactions, can then be regarded as the low energy effective operators
generated  by the three diagrams in Fig.~10, respectively.
\begin{figure}[t]
\centerline{\psfig{file=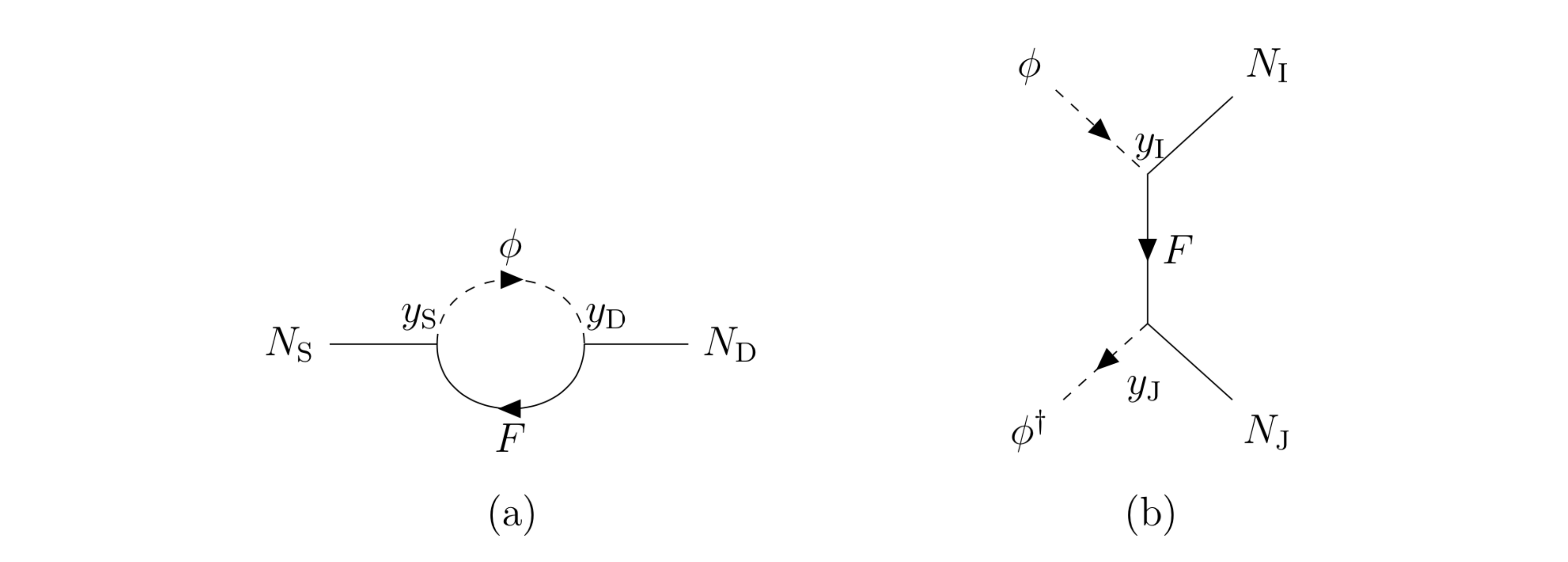,height=7cm,width=19cm,angle=0}}\vspace{0mm}
\caption{Feynman diagrams with a heavy fermion $F$ as mediator and $I,J={\rm D,S}$. Integrating out $F$, the diagram in the left panel
leads to the Feynman diagrams in panel (b) of Fig.~1.}
\end{figure}

If we take $M_{\rm F} \sim M_{\rm GUT}$, $y_{\rm S} \sim 1$ and $y_{\rm D} \sim 10^{-7}$, one can this time immediately 
reproduce the values $\widetilde{\L}_{\rm SS} \sim 10^{16}\,{\rm GeV}$,  $\widetilde{\L}_{\rm DS} \sim 10^{23}\,{\rm GeV}$ 
and $\widetilde{\L}_{\rm DD} \sim 10^{30}\,{\rm GeV}$: the correct values to reproduce the 
observed DM abundance from Higgs-induced RH neutrino mixing, with source neutrino Higgs portal interactions able to thermalise the source neutrino abundance prior to the onset of the oscillations and with a  suppressed contribution to dark neutrino production that we have indeed neglected. 
Notice that since the three couplings $\lambda_{IJ}$ are the product of just two Yukawa couplings, it is non trivial that the third comes out automatically satisfying correctly the condition for Higgs portal interactions producing dark neutrinos to be negligible. Moreover, 
they can be well understood imposing a ${\mathbb Z}_2$ symmetry under which all particles are even, except
the dark neutrino that is odd. In this way the small Yukawa coupling $y_{\rm D} \sim 10^{-7}$ could be regarded as a small symmetry 
breaking parameter connecting the visible sector to the dark sector.

\section{Summary}

The DM puzzle might have a solution at much higher scales than usually considered. Neutrino physics is a good place where to look for such a solution. 
A high scale RH neutrino playing the role of DM particle requires an extension of the SM beyond the minimal type-I seesaw Lagrangian (able to explain neutrino masses and mixing and the matter-antimatter asymmetry of the universe with leptogenesis). Higgs induced sterile-sterile neutrino mixing provides not only a way to produce the dark neutrinos with the right abundance but it also makes them detectable at neutrino telescopes. Higgs portal interactions for the source neutrino enhance the dark neutrino production and allow to lift the scale of leptogenesis up to 100 TeV. Intriguingly, the IceCube collaboration find an excess in the neutrino flux at $\sim$ 100 TeV that does not seem easy to explain with astrophysical sources but that can be well explained by RHINO DM decays. Further support omes from multimessenger consideration while a crucial test might come soon by an analysis of anisotropies in the 
very high energy neutrino flux at IceCube: stay tuned! 

\vspace{-1mm}
\subsection*{Acknowledgments}

I acknowledge financial support from the STFC Consolidated Grant ST/T000775/1.
I wish to thank Kareem Farrag, Patrick Ludl, Adam Murphy, 
Sergio Palomares-Ruiz, Rome Samanta and Ye-Ling Zhou for a fruitful  
collaboration on the RHINO DM model.

\end{document}